\documentclass[12pt]{article}

\usepackage{graphicx}
\usepackage{float}
\usepackage{cite}
\usepackage{amsfonts}
\usepackage{amssymb}
\usepackage{amsmath}
\usepackage{relsize}
\usepackage[compatibility=false]{caption}
\usepackage{subcaption}
\usepackage{xcolor}

\def\gtwid{\mathrel{\raise.3ex\hbox{$>$\kern-.75em\lower1ex\hbox{$\sim$}}}}
\def\ltwid{\mathrel{\raise.3ex\hbox{$<$\kern-.75em\lower1ex\hbox{$\sim$}}}}
\def\square{\kern1pt\vbox{\hrule height 1.2pt\hbox{\vrule width 1.2pt\hskip 3pt
   \vbox{\vskip 6pt}\hskip 3pt\vrule width 0.6pt}\hrule height 0.6pt}\kern1pt}

\begin{document}

\begin{titlepage}

\begin{flushright}
UFIFT-QG-24-02
\end{flushright}

\vskip 2.5cm

\begin{center}
{\bf The Price of Abandoning Dark Matter Is Nonlocality}
\end{center}

\vskip 1cm

\begin{center}
C. Deffayet$^{1*}$ and R. P. Woodard$^{2\dagger}$
\end{center}

\begin{center}
\it{$^{1}$ Laboratoire de Physique de l'Ecole Normale Sup\'{e}rieure,
ENS, Universit\'{e} PSL, CNRS
Sorbonne Universit\'{e}, Universit\'{e} Paris Cit\'{e}, F-75005 Paris, FRANCE}
\end{center}

\begin{center}
\it{$^{2}$ Department of Physics, University of Florida,\\
Gainesville, FL 32611, UNITED STATES}
\end{center}

\vspace{1cm}

\begin{center}
ABSTRACT
\end{center}
We consider the classic question posed by Pardo and Spergel about the
price of abandoning dark matter in the context of an invariant, 
metric-based theory of gravity. Our answer is that the price is 
nonlocality. This has been known for some time in the context of the 
quasi-static regime. We show that it also applies for cosmology and
we exhibit a model which reproduces standard CDM successes such as
perturbations in the cosmic microwave background, baryon acoustic
oscillations and structure formation. 

\begin{flushleft}
\end{flushleft}

\vspace{2cm}

\begin{flushleft}
$^{*}$ e-mail: cedric.deffayet@phys.ens.fr \\
$^{\dagger}$ e-mail: woodard@phys.ufl.edu
\end{flushleft}

\end{titlepage}

\section{Introduction}

Despite increasingly aggressive efforts at direct detection
\cite{PandaX-II:2017hlx,XENON:2018voc,PandaX-4T:2021bab,ADMX:2021nhd,
LZ:2022lsv,PerezAdan:2023rsl,XENON:2023cxc}, the evidence for dark 
matter is still restricted to its impact on gravity. It is therefore 
worthwhile considering whether or not the same gravitational 
phenomena might be instead signal some modification of gravity.
Milgrom's MOdified Newtonian Dynamics (MOND) \cite{Milgrom:1983ca,
Milgrom:1983pn,Milgrom:2014} does an excellent job of explaining, 
and even predicting observed features of galactic structure which 
are usually attributed to dark matter \cite{Milgrom:1983zz,
Milgrom:1989pr,Brada:1998mr,Brada:1998mi,Milgrom:2009bi,
McGaugh:2013zqa,Milgrom:2012xw,Lelli:2016uea,Milgrom:2016ogb,
McGaugh:2016leg,Lelli:2016cui,McGaugh:2020ppt,Banik:2021woo}.
MOND has mixed success explaining galactic clusters \cite{Aguirre:2001fj,
Milgrom:2018tqf,Clowe:2006eq}, but its real challenge is to reproduce the 
successes of dark matter in cosmology, including the anisotropies in the 
cosmic ray microwave background (CMB), baryon acoustic oscillations (BAO)
and linearized structure formation.

If dark matter is abandoned in favor of modified gravity, it must
be that MOND represents the static limit of some larger, relativistic
theory. So it is perhaps not surprising that extrapolations of MOND 
should encounter problems as one approaches the time-dependent regime
of cosmology. And it is significant that relativistic extensions of
MOND such as Bekenstein's TeVeS \cite{Bekenstein:2004ne} do a much 
better job of reproducing cold dark matter (CDM) cosmology 
\cite{Skordis:2005xk,Angus:2006qy,Dodelson:2006zt,Clifton:2011jh}. In 
particular, the AeST model of Skordis and Zlosnik \cite{Skordis:2020eui} 
is internally consistent and agrees with those phenomenological checks 
which have so far been made \cite{Skordis:2021mry,Verwayen:2023sds,
Bataki:2023uuy,Durakovic:2023out}.

The only metric-based modification of gravity which is both stable and
generally coordinate invariant consists of changing the Einstein-Hilbert
Lagrangian from $R$ to some nonlinear function $f(R)$ \cite{Woodard:2006nt}.
This induces a new scalar degree of freedom which could be regarded as a 
sort of dark matter \cite{Cembranos:2008gj}. Within the context of locality,
the other options for modifying gravity involve either abandoning the metric 
as the sole gravitational field variable --- such as TeVeS 
\cite{Bekenstein:2004ne} and AeST \cite{Skordis:2020eui} --- or else 
abandoning full general coordinate invariance \cite{Horava:2009uw,
Barvinsky:2015kil}. We will here explore the option of abandoning locality.

Our result can be regarded as an answer to the question posed in the title
of the recent paper by Pardo and Spergel, ``What is the price of abandoning 
dark matter?'' \cite{Pardo:2020epc}. We demonstrate that the price is 
nonlocality, provided one maintains both general coordinate invariance and 
the metric's status as the sole gravitational field. This has already been 
shown for gravitationally bound systems \cite{Deffayet:2011sk,
Deffayet:2014lba}, but cosmological extensions of that model involve 
fitting a free function just to recover the correct expansion history 
\cite{Kim:2016nnd}, and in any case fail to provide the extra gravitational
force need for structure formation without dark matter \cite{Tan:2018bfp}.
In this new effort we exploit the equations which describe cold dark matter
(CDM) cosmology in the regime of linearized perturbations, in particular, 
the fact that there is no strong evidence for self-interactions
\cite{Tulin:2017ara,Sarkar:2018xon,Adhikari:2022sbh}. This means that the
CDM stress tensor must be separately conserved, which permits us to express
it in terms of linearized scalar gravitational perturbations. We accomplish 
this in section 2, and also infer a gravitational effective action whose 
variation gives the CDM stress tensor. In section 3 we devise a generally 
coordinate invariant expression for this effective action. This approach is
guaranteed to reproduce the cosmology of CDM. In our Conclusion (section 4) 
we discuss the possibility that this action arises from secular interactions 
between inflationary gravitons which became nonperturbatively strong
\cite{Woodard:2018gfj}. We also discuss the prospects for a single effective 
action which describes both cosmology and gravitationally bound systems. 

\section{Reconstructing the Effective Action}

The purpose of this section is to derive a purely gravitational effective
action whose variation is guaranteed to reproduce CDM phenomenology in
the linearized regime. We begin by giving the metric perturbations and
the two variables which characterize the CDM stress tensor. The fact that
the CDM stress tensor is separately conserved permits us to express the
density contrast and momentum divergence in terms of the metric. The
section closes by constructing a gravitational effective action whose  
variation gives the CDM stress tensor.

\subsection{Linearized Perturbations}

It is simple to reproduce the CDM contribution to the expansion history 
because this is just a single degree of freedom. The real challenge is 
getting the infinite number of degrees of freedom characterized by 
linearized perturbations. Although there are four scalar perturbations
about the Friedmann-Lema\^itre-Robertson-Walker (FLRW) geometry, gauge
transformations can be used to set any two to zero. We choose the two 
nonzero perturbations as,
\begin{equation}
ds^2 = -\Bigl[ 1 + 2 \Psi(t,\vec{x}) \Bigr] dt^2 + 2 \partial_{i} 
B(t,\vec{x}) dt dx^i + a^2(t) d\vec{x} \cdot d\vec{x} \; . \label{geometry}
\end{equation}
The first order connections are,
\newpage 
\begin{eqnarray}
\Gamma^{0}_{~00} &\!\!\! = \!\!\!& \dot{\Psi} \;\; , \;\; \Gamma^{0}_{~0i} = 
\partial_i \Bigl[\Psi + H B\Bigr] \;\; , \;\; \Gamma^{0}_{~ij} = H a^2 
\delta_{ij} \Bigl[1 \!-\! 2 \Psi \Bigr] - \partial_{i} \partial_{j} B \; , 
\qquad \label{Gamma1} \\
\Gamma^{i}_{~00} &\!\!\! = \!\!\!& \frac{\partial_i}{a^2} \Bigl[ \Psi \!+\! 
\dot{B}\Bigr] \qquad , \qquad \Gamma^{i}_{~j0} = H \delta_{ij} \qquad , \qquad 
\Gamma^{i}_{~jk} = -H \partial_{i} B \delta_{jk} \; . \label{Gamma2} \qquad 
\end{eqnarray}
The CDM stress tensor $\Delta T_{\mu\nu}$ is characterized (at linearized
order) by its background energy density $\overline{\rho}(t)$, the density 
contrast $\delta(t,\vec{x}) \equiv \delta \rho(t,\vec{x}) \div 
\overline{\rho}(t)$ and the momentum divergence $\theta(t,\vec{x})$,
\begin{equation}
\Delta T_{00} = \overline{\rho} \Bigl[1 + \delta + 2 \Psi + \ldots\Bigr] \; , \; 
\Delta T_{0i} = \overline{\rho} \partial_{i} \Bigl[0 - \theta + \ldots\Bigr]
\;\; , \;\; \Delta T_{ij} = \overline{\rho} \Bigl[0 + 0 + \ldots\Bigr] \; .
\label{stress}
\end{equation}

\subsection{Eliminating CDM}

We now exploit the fact that the CDM stress tensor is separately 
conserved. Relations (\ref{Gamma1}-\ref{Gamma2}) can be used to derive  
two equations,
\begin{eqnarray}
0 = D^{\mu} \Delta T_{\mu 0} & \!\!\! = \!\!\! & -(\partial_t \!+\! 3 H) 
\overline{\rho} + \overline{\rho} \Bigl\{ -\dot{\delta} \!+\! 
\frac{\nabla^2}{a^2} \Bigl[B \!-\! \theta\Bigr] \Bigr\} + \dots \; , 
\qquad \label{cons0} \\
0 = D^{\mu} \Delta T_{\mu i} & \!\!\! = \!\!\! & 0 + \overline{\rho} 
\partial_{i} \Bigl\{\dot{\theta} \!+\! \Psi\Bigr\} + \dots \; . 
\label{consi}
\end{eqnarray}
The 0th order part of equation (\ref{cons0}) implies that the background
CDM energy density has the equation of state of pressureless matter,
\begin{equation}
\overline{\rho}(t) = \frac{\rho_0}{a^3(t)} \; , \label{rhobar}
\end{equation}
where $\rho_0$ is a constant. Equation (\ref{consi}) allows us to express
the momentum divergence as the time integral of the Newtonian potential,
\begin{equation}
\theta(t,\vec{x}) = -\frac{1}{\partial_t} \Psi(t,\vec{x}) \equiv -\int_{0}^{t}
\!\! dt' \, \Psi(t',\vec{x}) \; . \label{theta}
\end{equation}
Substituting (\ref{theta}) in equation (\ref{cons0}) gives the density 
contrast,
\begin{equation}
\delta = \frac{1}{\partial_t} \Bigl[ \frac{\nabla^2}{a^2} \Bigl( 
B \!+\! \frac{1}{\partial_t} \Psi\Bigr) \Bigr] \; . \label{delta}
\end{equation}

\subsection{Reconstructing the Effective Action}

We seek an effective action $\Delta S$ whose variation gives the CDM
stress tensor (\ref{stress}), considered as a functional of the metric.
The general relation between an induced stress tensor and variations is,
\newpage
\begin{equation}
\Delta T_{\mu\nu} = -\frac{2}{\sqrt{-g}} \frac{\delta \Delta S}{\delta
g^{\mu\nu}} \qquad \Longleftrightarrow \qquad \frac{\delta \Delta S}{\delta
g^{\mu\nu}} = -\frac12 \sqrt{-g} \, \Delta T_{\mu\nu} \; . \label{DStoT}
\end{equation}
Because our metric (\ref{geometry}) depends only on the two scalar 
perturbations $\Psi(t,\vec{x})$ and $B(t,\vec{x})$ we must specialize 
relation (\ref{DStoT}) to variations with respect to these two fields,
\begin{eqnarray}
\frac{\delta \Delta S}{\delta \Psi} &\!\!\! = \!\!\!& -2 \frac{\partial
g^{\mu\nu}}{\partial g_{00}} \!\times\! \frac{\partial \Delta S}{\delta
g^{\mu\nu}} = -\sqrt{-g} \, g^{0\mu} g^{0\nu} \Delta T_{\mu\nu} \; , 
\qquad \label{varPsi} \\
\frac{\delta \Delta S}{\delta B} &\!\!\! = \!\!\!& -2 \partial_{i} 
\Bigl\{ \frac{\partial g^{\mu\nu}}{\partial g_{0i}} \!\times\! 
\frac{\partial \Delta S}{\delta g^{\mu\nu}} \Bigr\} = -\partial_{i}
\Bigl\{ \sqrt{-g} \, g^{0\mu} g^{i\nu} \Delta T_{\mu\nu} \Bigr\} \; . 
\qquad \label{varB} 
\end{eqnarray} 
Substituting (\ref{geometry}) and (\ref{stress}), and retaining only
terms up to first order gives,
\begin{equation}
\frac{\delta \Delta S}{\delta \Psi} = -a^3 \overline{\rho} \Bigl\{1 \!-\! 
\Psi \!+\! \delta \!+\! \ldots\Bigr\} \;\; , \;\; \frac{\delta \Delta S}{
\delta B} = -a^3 \overline{\rho} \frac{\nabla^2}{a^2} \Bigl\{-B \!+\! 
\theta \!+\! \ldots\Bigr\} \; . \label{varsfinal}
\end{equation}

There are two ways to reconstruct an effective Lagrangian $\Delta 
\mathcal{L}$ whose variations reproduce (\ref{varsfinal}). The first is
to consider $\delta(t,\vec{x})$ and $\theta(t,\vec{x})$ as auxiliary
fields whose equations of motion enforce relations 
(\ref{theta}-\ref{delta}). In that case the result is,
\begin{equation}
\Delta \mathcal{L}_1 = -\rho_0 \Bigl\{ 1 + \Psi - \frac12 \Psi^2 -
\Bigl[ B - \theta\Bigr] \frac{\nabla^2}{2 a^2} \Bigl[ B - \theta \Bigr] 
+ \delta (\dot{\theta} + \Psi) + \ldots \Bigr\} . \label{L1}
\end{equation}
Note that we have used $\overline{\rho}(t) \times a^3(t) = \rho_0$, 
and that the variation of (\ref{L1}) with respect to $\delta$ 
enforces (\ref{theta}) while the variation with respect to $\theta$
gives (\ref{delta}). 

The second way to realize (\ref{varsfinal}) is through a nonlocal 
Lagrangian which depends only on $\Psi$ and $B$,
\begin{equation}
\Delta \mathcal{L}_2 = -\rho_0 \Bigl\{ 1 + \Psi - \frac12 \Psi^2 -
\Bigl[B +  \frac{1}{\partial_t} \Psi \Bigr] \frac{\nabla^2}{2 a^2} 
\Bigl[ B + \frac{1}{\partial_t} \Psi \Bigr] + \ldots \Bigr\} . \label{L2}
\end{equation}
In this case one must interpret the variation of $\frac1{\partial_t} 
\Psi$ in the sense of nonlocal cosmology \cite{Deser:2007jk}. That 
means replacing the ``advanced'' integral, which is what really results 
from the variation, with the ``retarded'' integral which obeys the 
same differential equation,
\begin{equation}
\int_{t}^{\infty} \!\!\!\! dt' F(t') \longrightarrow -\int_{0}^{t} \!\! 
dt' F(t') \; . \label{oldtrick}
\end{equation}
The actual genesis of a nonlocal model would be through the 
Schwinger-Keldysh formalism \cite{Schwinger:1960qe,Mahanthappa:1962ex,
Bakshi:1962dv,Bakshi:1963bn,Keldysh:1964ud,Chou:1984es}, with 
(\ref{oldtrick}) rigorously justified through the interference between 
contributions from forward and backwards propagating effective fields 
\cite{Jordan:1986ug,Calzetta:1986ey,Ford:2004wc}.

\section{Invariant Formulation}

The purpose of this section is to construct an invariant effective 
action whose variation produces the full CDM stress tensor. We begin
with a convenient representation for the general metric. Then it is
shown how the full CDM stress tensor can be expressed in terms of
two scalar fields which obey simple, first order differential equations.
The section closes by giving an invariant Lagrangian whose variation
reproduces the full CDM stress tensor.

\subsection{The ADM Metric}

Because of the crucial role played by time evolution, it is 
convenient to carry out this analysis using the representation for the 
full metric introduced by Arnowitt, Deser and Misner \cite{Arnowitt:1959ah},
\begin{equation}
ds^2 = -N^2 dt^2 + \gamma_{ij} \Bigl( dx^i - N^i dt\Bigr) \Bigl(
dx^j - N^j dt\Bigr) \; . \label{ADM}
\end{equation}
The various components are known as the lapse $N(t,\vec{x})$, the shift
$N^i(t,\vec{x})$ and the 3-metric $\gamma_{ij}(t,\vec{x})$. In these 
variables the measure and the inverse metric are,
\begin{equation}
\sqrt{-g} = N \sqrt{\gamma} \qquad , \qquad g^{\mu\nu} = 
\left( \begin{matrix} 
-\frac{1}{N^2} & -\frac{N^n}{N^2} \\
-\frac{N^m}{N^2} & \gamma^{mn} - \frac{N^m N^n}{N^2} \\
\end{matrix} \right) \; . \label{detinv}
\end{equation}
For the scalar perturbation geometry (\ref{geometry}) we have,
\begin{equation}
N^2 = 1 + 2 \Psi + \frac{\partial_i B}{a} \frac{\partial_i B}{a} 
\qquad , \qquad N^i = -\frac{\partial_i B}{a^2} \qquad , \qquad
\gamma_{ij} = a^2 \delta_{ij} \; . \label{scalarADM}
\end{equation}

\subsection{Representing the Full CDM Stress Tensor}

Because cold dark matter has zero pressure and is collisionless, its stress
tensor can be written in terms of an energy density $\rho(t,\vec{x})$ and a 
timelike 4-velocity field $u_{\mu}(t,\vec{x})$
\begin{equation}
T_{\mu\nu} = \rho \, u_{\mu} u_{\nu} \qquad , \qquad g^{\mu\nu} u_{\mu}
u_{\nu} = -1 \; . \label{genform}
\end{equation}
Comparison with (\ref{stress}) and (\ref{theta}) implies an intriguing 
expansion for the 4-velocity,
\begin{eqnarray}
u_0 &\!\!\! = \!\!\!& -1 - \Psi + \ldots = -\partial_{t} \Bigl[t + 
\frac1{\partial_t} \Psi + \ldots\Bigr] \; , \qquad \label{u0} \\
u_i &\!\!\! = \!\!\!& 0 + \partial_i \theta + \ldots = -\partial_i \Bigl[
t + \frac1{\partial_t} \Psi + \ldots\Bigr] \; . \qquad \label{ui}
\end{eqnarray}
This suggests that $u_{\nu}$ can be written as minus the gradient of a
scalar whose expansion resides within the square brackets.

We can understand the expansions (\ref{u0}-\ref{ui}) using the fact that 
the CDM stress tensor is separately conserved,
\begin{equation}
D_{\mu} T^{\mu}_{~\nu} = D_{\mu} \Bigl( u^{\mu} \rho\Bigr) u_{\nu}
+ \rho \, u^{\mu} D_{\mu} u_{\nu} = 0 \; . \label{gencons}
\end{equation}
Conservation implies two distinct equations,
\begin{eqnarray}
\partial_{\mu} \Bigl(\sqrt{-g} \, g^{\mu\nu}  u_{\nu} \rho \Bigr) & = & 0 \; , 
\label{cons1} \\
g^{\mu\rho} u_{\mu} D_{\rho} u_{\nu} & = & 0 \; . \label{cons2}
\end{eqnarray}
Relation (\ref{cons2}) can indeed be solved as $u_{\nu} = -\partial_{\nu} 
\phi$, where the scalar field $\phi(t,\vec{x})$ obeys the nonlinear, first 
order differential equation,
\begin{equation}
g^{\mu\nu} \partial_{\mu} \phi \partial_{\nu} \phi = -1 \; . \label{scalar}
\end{equation}
The correctness of this solution can be seen by substituting $u_{\nu} = 
-\partial_{\nu} \phi$ in relation (\ref{cons2}) and then using 
(\ref{scalar}),
\begin{equation}
g^{\mu\rho} u_{\mu} D_{\rho} u_{\nu} = g^{\mu\rho} \partial_{\mu} \phi
D_{\rho} \partial_{\nu} \phi = g^{\mu\rho} \partial_{\mu} \phi D_{\nu}
\partial_{\rho} \phi = \frac12 D_{\nu} \Bigl[ g^{\mu\rho} \partial_{\mu} 
\phi \partial_{\rho} \phi\Bigr] = 0 \; . \label{proof}
\end{equation}

Expressions (\ref{scalar}) and (\ref{cons1}) give well-posed evolution
equations for $\phi(t,\vec{x})$ and $\rho(t,\vec{x})$, respectively. 
To see this for $\phi(t,\vec{x})$, substitute (\ref{ADM}) into 
(\ref{scalar}) and then solve for $\dot{\phi}$ with an irrelevant sign 
choice,
\begin{equation}
\dot{\phi} = N \sqrt{1 + \gamma^{ij} \partial_i \phi \partial_j \phi}
- N^i \partial_i \phi \; . \label{evophi}
\end{equation}
Equation (\ref{evophi}) obviously has a unique solution which will agree
with our weak field expansions (\ref{u0}-\ref{ui}) if we choose the 
initial condition $\phi(0,\vec{x}) = 0$. Substituting (\ref{ADM}) into 
(\ref{cons1}) and then eliminating $\dot{\phi}$ using (\ref{evophi}) 
gives a similarly well-posed evolution equation for $\rho(t,\vec{x})$,
\begin{equation}
\partial_{t} \Bigl[ \rho \sqrt{\gamma} \, \sqrt{1 \!+\! \gamma^{jk}
\partial_j \phi \partial_k \phi} \, \Bigr] = \partial_{i} \Bigl[N \rho
\sqrt{\gamma} \, \gamma^{ij} \partial_j \phi - N^i \rho \sqrt{\gamma} \, 
\sqrt{1 \!+\! \gamma^{jk} \partial_j \phi \partial_k \phi} \, \Bigr] 
\; . \label{evorho}
\end{equation}
This also has a unique solution which will agree with (\ref{delta}) if 
we begin with,
\begin{equation}
\rho(0,\vec{x}) = \frac{\rho_0}{\sqrt{\gamma(0,\vec{x})}} \; . 
\label{initialrho}
\end{equation}
Note finally that the method of characteristics gives a solution for 
$\rho(t,\vec{x})$ in terms of $\phi(t,\vec{x})$ and 
$g_{\mu\nu}(t,\vec{x})$.

\subsection{Nonlocal Effective Action}

We can reproduce the full CDM stress tensor (\ref{genform}), for a 
general metric, using a local effective Lagrangian with auxiliary fields 
$\rho(t,\vec{x})$ and $\phi(t,\vec{x})$,
\begin{equation}
\mathcal{L}_1 = -\frac12 \Bigl[\partial_{\mu} \phi \partial_{\nu} 
\phi g^{\mu\nu} + 1 \Bigr] \rho \sqrt{-g} \; . \label{fullL1}
\end{equation}
The variation with respect to $\phi$ gives the $\rho$ equation 
(\ref{cons1}), while the variation with respect to $\rho$ implies the
$\phi$ equation (\ref{scalar}),
\begin{eqnarray}
\frac{\delta S_1}{\delta \phi} &\!\!\! = \!\!\!& \partial_{\mu} \Bigl[ 
\rho \sqrt{-g} \, g^{\mu\nu} \partial_{\nu}
\phi \Bigr] = 0 \; , \label{varphi} \\
\frac{\delta S_1}{\delta \rho} &\!\!\! = \!\!\!& -\frac12 \Bigl[ 
\partial_{\mu} \phi \partial_{\nu} \phi g^{\mu\nu} + 1 \Bigr] \sqrt{-g}
= 0 \; . \label{varrho}
\end{eqnarray}
The variation with respect to the metric reproduces the full CDM stress
tensor (\ref{genform}) when equation (\ref{varrho}) is used and we recall
that the 4-velocity is $u_{\mu} = -\partial_{\mu} \phi$,
\begin{equation}
T_{\mu\nu} \equiv -\frac{2}{\sqrt{-g}} \frac{\delta S_1}{\delta g^{\mu\nu}}
= \rho \, \partial_{\mu} \phi \partial_{\nu} \phi = \rho \, u_{\mu} u_{\nu} 
\; . \label{vargmn}
\end{equation}

Lagrangian (\ref{fullL1}), localized in terms of auxiliary scalars $\phi$
and $\rho$, is analogous to the weak field result (\ref{L1}) which was
localized in terms of auxiliary scalars $\theta$ and $\delta$. We stress
that the fields $\phi(t,\vec{x})$ and $\rho(t,\vec{x})$ in the CDM stress
tensor (\ref{vargmn}) are not independent dynamical variables but rather 
nonlocal functionals of the metric which solve the well-posed evolution 
equations (\ref{evophi}) and (\ref{evorho}), with initial value data
$\phi(0,\vec{x}) = 0$ and (\ref{initialrho}), respectively. It would be 
possible to derive a purely gravitational effective action, analogous to
the weak field expression (\ref{L2}), as a surface term. However, it is
really equation (\ref{vargmn}) which defines the model.

\section{Conclusions}

We have derived a nonlocal, metric-based effective action which exactly
reproduces cold dark matter. For linearized scalar perturbations in 
cosmology (\ref{geometry}) our result can be expressed either as a localized 
form (\ref{L1}), using auxiliary scalars $\theta$ and $\delta$, or else as 
a nonlocal, purely gravitational form (\ref{L2}). Either way, the effective 
CDM stress tensor takes the form (\ref{stress}), with (\ref{theta}-\ref{delta}).
For a general metric (\ref{ADM}) the localized form is (\ref{fullL1}), with
auxiliary scalars $\phi$ and $\rho$. The nonlocal, purely gravitational form
analogous to (\ref{L2}) would be a surface term. Either way, the effective
CDM stress tensor takes the form (\ref{vargmn}), with the auxiliary scalars
defined as nonlocal functionals for the metric which obey the well-posed
evolution equations (\ref{evophi}) and (\ref{evorho}), starting from initial
values $\phi(0,\vec{x}) = 0$ and (\ref{initialrho}). 

We stress that there can be no doubt about this model reproducing all the
successes of CDM in cosmology. Those successes include the anisotropies of 
the cosmic microwave background, baryon acoustic oscillations, and linearized 
structure formation. All of this must come out right because this model was 
constructed by using the separate conservation of the CDM stress tensor to
express it as a nonlocal functional of the metric. The only way this model
can be falsified is by showing that CDM interacts with fields other than 
gravity, the evidence for which is weak \cite{Tulin:2017ara,Sarkar:2018xon,
Adhikari:2022sbh}. This model should put to rest the frequent claims that no 
modified gravity theory can supplant dark matter. It also demonstrates that 
the answer to the question of Pardo and Spergel \cite{Pardo:2020epc} about 
the price of abandoning dark matter: the price is nonlocality.

A key principle in constructing modified gravity models is that 
gravitational radiation in vacuum should move at the speed of light so as
not to conflict with the simultaneous detection of optical and gravitational
radiation from the binary neutron star merger GW170817 approximately 40 Mpc 
distant \cite{Boran:2017rdn}. This is usually accomplished by making each 
term in the effective action involve at least two factors of the Ricci 
tensor. We did not need to do that here because our model exactly reproduces 
the result of the CDM stress tensor. So any change our model predicts in the
kinematics of gravitational radiation is limited to one already predicted by
cold dark matter.

One thing our model does not do is to agree with the nonlocal extension of
MOND \cite{Deffayet:2011sk,Deffayet:2014lba} which was constructed to 
reproduce the baryonic Tully-Fisher relation \cite{McGaugh:2000sr}, with
sufficient weak lensing, in gravitationally bound structure. One can see 
this by simply counting derivatives and weak fields in the CDM stress tensor.
At least in cosmological perturbation theory, our model's change in the 
gravitational field equations is linear in the weak fields $\Psi(t,\vec{x})$ 
and $B(t,\vec{x})$, and every inverse differential operator is compensated 
by a positive power. In contrast, the model which works inside gravitationally 
bound structures is quadratic in the weak fields and involves three 
derivatives \cite{Deffayet:2011sk,Deffayet:2014lba}. It would be desirable
to have a single formalism which connects both regimes.

It would also be desirable to derive these nonlocal models from fundamental 
theory. We believe they might arise from resumming the secular logarithms 
that are induced by loops of inflationary gravitons and which must 
eventually become nonperturbatively large during a prolonged period of
inflation \cite{Woodard:2018gfj}. Such a resummation has been accomplished
for the similar logarithms produced by nonlinear sigma models on de Sitter
background \cite{Miao:2021gic}. The curvature-induced effective potentials
which enable the resummation have been generalized from de Sitter to an 
arbitrary cosmological background which experiences primordial inflation 
\cite{Kasdagli:2023nzj}. And when the generalized effective potentials are
used to evolve to late times, certain results do retain a strong memory of
the inflationary epoch \cite{Woodard:2023cqi}. It remains to generalize 
the resummation technique from nonlinear sigma models to gravity.

Finally, one should keep in mind the possibility that MOND might be 
realized as a modification of inertia rather than as a modification of
gravity \cite{Milgrom:2023pmv}. In this regard it is worth noting that
inflationary gravitons not only induce secular changes in the force of 
gravity \cite{Tan:2022xpn}, they also modify the kinematics (and hence
the inertia) of massless fermions \cite{Miao:2006gj} and photons 
\cite{Wang:2014tza}. It would be interesting to explore changes to the
inertia, and particularly the response to gravity, of massive particles
\cite{Miao:2012bj}.

\vskip .5cm

\centerline{\bf Acknowledgements}

We are grateful to C. Skordis for patiently explaining his model and
the associated constraints of CDM in cosmology. We are also grateful
for discussions and correspondence with M. Milgrom and S. Sarkar. This 
work was partially supported by NSF grant PHY-2207514 and by the 
Institute for Fundamental Theory at the University of Florida.

\end{document}